\documentclass[preprint]{aastex}

\usepackage{amstext}
\usepackage{amssymb}
\usepackage{graphicx}
\usepackage{xspace}
\usepackage{enumitem}

\newcommand{\JWST}{\textit{JWST}\xspace}
\newcommand{\WFIRST}{\textit{WFIRST}\xspace}
\newcommand{\IRSSquare}{IRS$^2$\xspace}

\newcommand{\ie}{\textit{i.e.}\xspace}

\slugcomment{\small Accepted by Publications of the Astronomical Society of the Pacific}

\begin{document}

\title{Teledyne H1RG, H2RG, and H4RG Noise Generator}

\author{Bernard J. Rauscher}
\affil{NASA Goddard Space Flight Center, Greenbelt, MD, 20771, USA}
\email{Bernard.J.Rauscher@nasa.gov}

\begin{abstract}
This paper describes the near-infrared detector system noise generator (NG) that we wrote for the James Webb Space Telescope (\JWST) Near Infrared Spectrograph (NIRSpec). NG simulates many important noise components including; (1) white ``read noise'', (2) residual bias drifts, (3) pink $1/f$ noise, (4) alternating column noise, and (5) picture frame noise. By adjusting the input parameters, NG can simulate noise for Teledyne's H1RG, H2RG, and H4RG detectors with and without Teledyne's SIDECAR ASIC IR array controller. NG can be used as a starting point for simulating astronomical scenes by adding dark current, scattered light, and astronomical sources into the results from NG. NG is written in Python-3.4. The source code is freely available for download from http://jwst.nasa.gov/publications.html.
\end{abstract}


\section{Introduction\label{Sec:Intro}}

Simulated detector noise is a standard tool for observation planning. Unfortunately, the noise of real instruments can be complex. For example, it may contain correlations that appear as banding, alternating column noise (ACN), residual bias or ``pedestal'' drifts, and bias patterns that fade in and out. Simple linear combinations of read noise, shot noise, and even $1/f$ noise can fail to satisfactory reflect the full complexity of real data.

This article describes NG, a read noise simulator that we wrote for the \JWST NIRSpec detector subsystem. NG builds on a foundation of principal components analysis (PCA) and the read noise studies that were done to develop NIRSpec's Improved Reference Sampling and Subtraction (\IRSSquare; pronounced ``IRS-square'') readout mode \citep{Moseley:2010kc,Rauscher:2013ht}. NG faithfully reproduces most of the Fourier noise power seen in real NIRSpec darks, and by extension darks from other HxRG based systems. NG includes uncorrelated, correlated, stationary, and non-stationary noise components. It correctly accounts for end of row and end of frame timing gaps. By changing the input parameters, NG can generate read noise for other Teledyne HxRG family detectors including the H1RG, H4RG-10 (planned for WFIRST-AFTA\footnote{Wide Field Infrared Survey Telescope - Astrophysics Focused Telescope Assets}), and H4RG-15 (planned for ground based astronomy).

Although NG faithfully reproduces most of NIRSpec's detector subsystem noise, there are a few things that it leaves out. The current version does not simulate random telegraph noise \citep[RTN;][]{Rauscher:2004ga,Bacon:2004dq,Bacon:2005bl}. Furthermore, we treat the non-stationary picture frame as a small perturbation on the dominant stationary noise. A more rigorous treatment would allow for mixing of the non-stationary and stationary components. In spite of these limitations, we believe that NG is a useful tool that offers higher fidelity simulations than can be easily achieved otherwise. 

\section{Getting Started Quickly\label{sec:quick_start}}

NG is written in Python-3.4 and requires the astropy, datetime, numpy, os, scipy, and warnings modules to run. The README file specifies which versions of these modules we are running. We typically run NG using IPython Notebook. Here are the steps to install and run the examples on a unix computer that already has these components installed.

\begin{enumerate}
\item Download the distribution from http://jwst.nasa.gov/publications.html and put it somewhere in your Python search path.
\item Uncompress the archive using gunzip.
\item Unpack the tar file using tar.
\item cd into the nghxrg/ subdirectory.
\item Configure a shell environment variable named NGHXRG\_HOME to point to it.
\item Type ``ipython notebook'' to start IPython Notebook.
\item Select the file 01\_README.ipynb to start the examples.
\end{enumerate}

The file 02\_README.pdf contains the same information in the more widely used, but non-executable, PDF format. The examples include generating read noise for \JWST NIRSpec, H4RGs like those that are planned for \WFIRST, and a few examples that highlight specific noise components.

\section{NG's Physical Basis\label{sec:physics}}

\subsection{Appearance of Noise\label{sec:appearance}}

Although every near-infrared (NIR) detector system is different, there are usually some similarities. In this article, we take the \JWST NIRSpec detector subsystem as our proxy for a HxRG based system. Although the relative amplitudes of the noise components are specific to NIRSpec, in our experience the same general noise components are often seen in other contexts.

\begin{figure}[ht]\begin{center}
\includegraphics[width=6in]{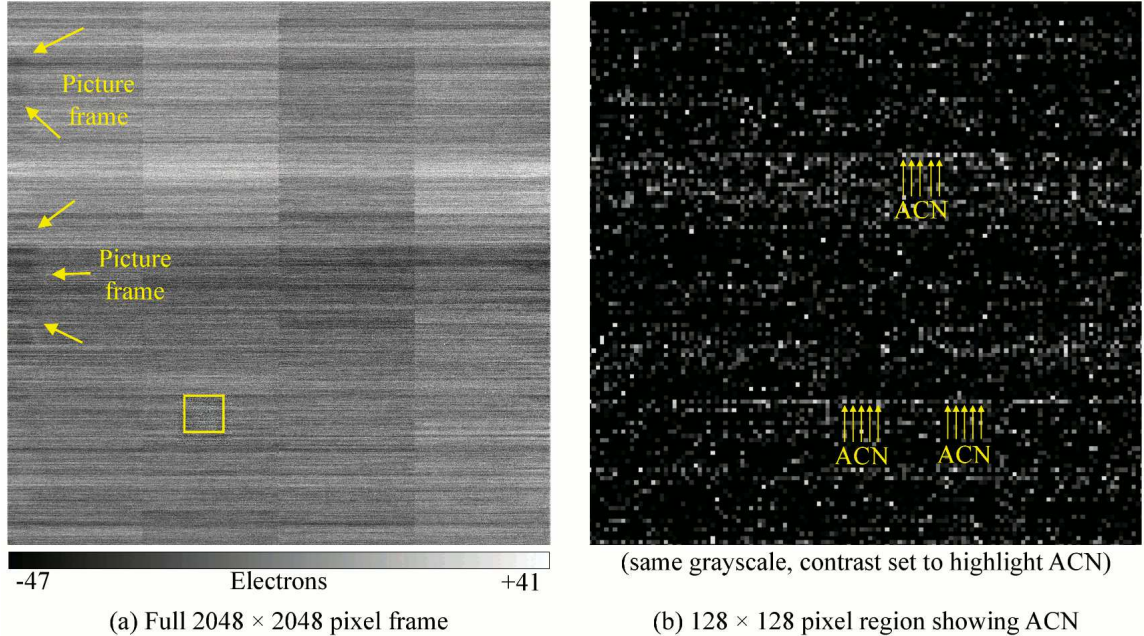}
\caption{(a) Each NIRSpec H2RG has $2048\times 2048$~pixels. This reference corrected full frame CDS dark is typical for NIRSpec's H2RGs when driven by the $\textrm{T}\sim 40$~K SIDECARs. The standard deviation is about $\sigma_{\rm read}=14~e^-$~rms. The four video outputs are visible as thick $512\times 2048$~pixel stripes. The horizontal banding is caused mostly by $\sim 1/f$ drifts in the bias voltages generated by the SIDECAR. The yellow arrows point to picture frame noise. There is more picture frame noise on the opposite side of this image that we did not highlight to avoid obscuring it. In (b), we expand the yellow $128\times 128$~pixel box to show ACN. Because the NIRSpec detector subsystem was tuned to minimize ACN, it is not easy to see.  One of the downloadable examples has been tuned to more clearly show ACN.\label{fig:noise_overview}}
\end{center}\end{figure}

Fig.~\ref{fig:noise_overview}a shows a typical reference corrected NIRSpec correlated double sampling (CDS) dark integration.\footnote{In this article, we use a simple reference correction scheme. We treat each output separately and subtract the median value of all reference pixels in rows from every pixel on that output.} Visual inspection reveals white read noise and pink noise. The pink noise manifests as horizontal banding. More detailed analysis reveals pedestal drifts, ACN (Fig.~\ref{fig:noise_overview}b), and a faint picture frame pattern that fades in and out. Pedestal drifts are residual bias offsets between integrations that are detectable even after dark subtraction. To a good approximation, all pixels in the detector move together in a pedestal drift. The examples (see Sec.~\ref{sec:implementation}) generate FITS files that show ACN and picture frame noise more clearly than is practical here.

\begin{figure}[ht]\begin{center}
\includegraphics[width=6in]{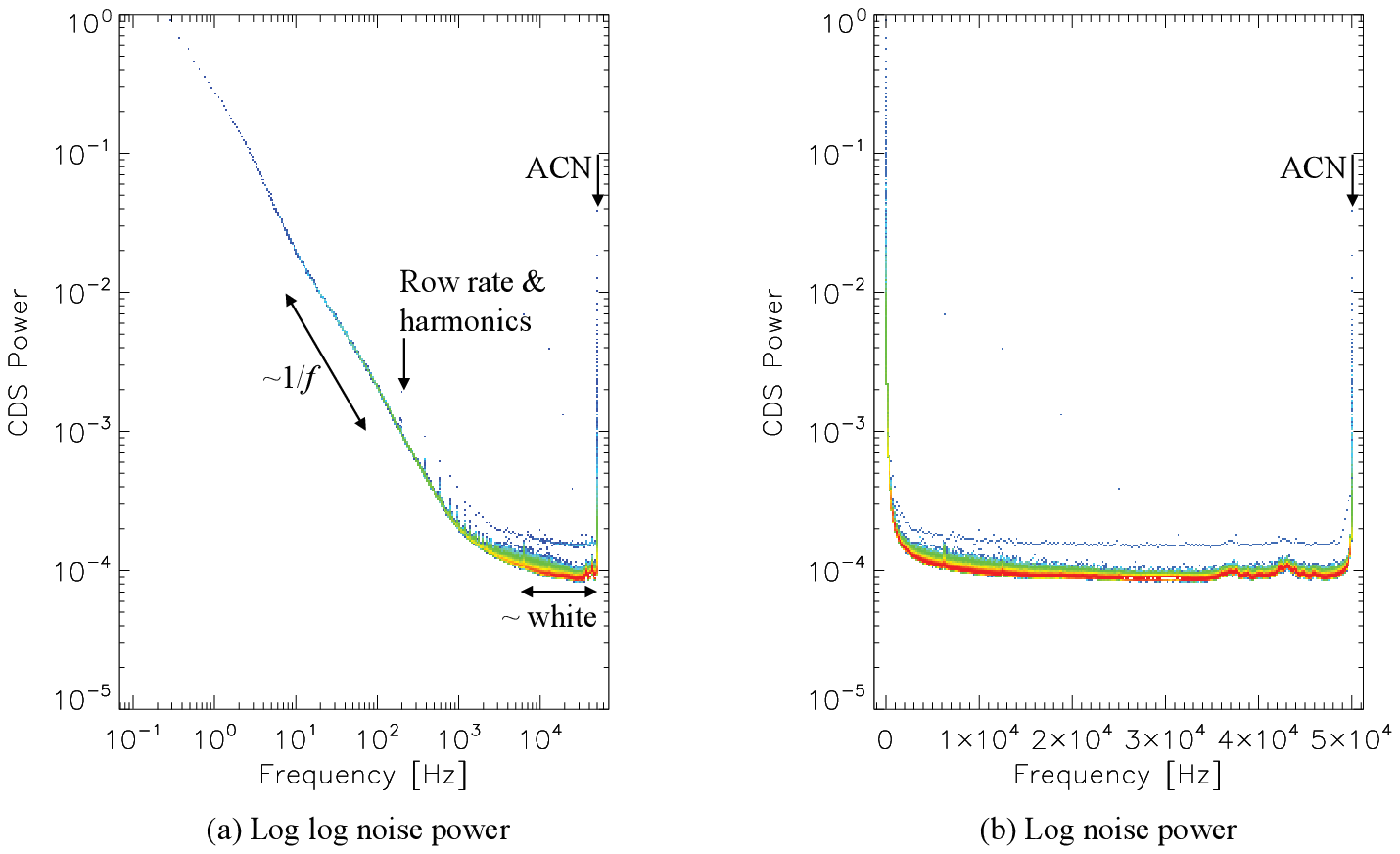}
\caption{Stationary noise is easiest to understand in Fourier space. In these heavily averaged  power spectra, the colors indicate the density of data points from (low density) blue to (high density) red. When displayed on an (a) log-log scale, the $\sim 1/f$ low frequency noise is obvious. In (b), we use a linear-log scale to better show ACN. ACN manifests as a strong feature at the Nyquist frequency, 50~kHz. The flat part of the power spectrum is roughly white. NG does not attempt to model all of the lines, bumps, and wiggles that are shown here. For example, we do not attempt to model the lines associated with the row rate and harmonics because these contain little power.
\label{fig:npsds}}
\end{center}\end{figure}

Most of NIRSpec's noise is stationary: meaning that the covariance matrix is the same at all times. The covariance matrix of stationary noise is diagonal in Fourier space. For this reason, the Fourier noise power spectrum (Fig.~\ref{fig:npsds}) provides a particularly powerful way of looking at NIRSpec's noise. Here the horizontal banding clearly appears as $1/f$ noise. The power spectrum also reveals a line at the Nyquist frequency with a $1/f$-like wing on the low frequency side. This is ACN. Although the flat part of the power spectrum at intermediate frequencies is not perfectly flat, in NG we treat it as if it were. This partially explains why NG does not account for all of the read noise.

\subsection{Noise Origins\label{sec:origins}}

\begin{figure}[ht]\begin{center}
\includegraphics[width=3in]{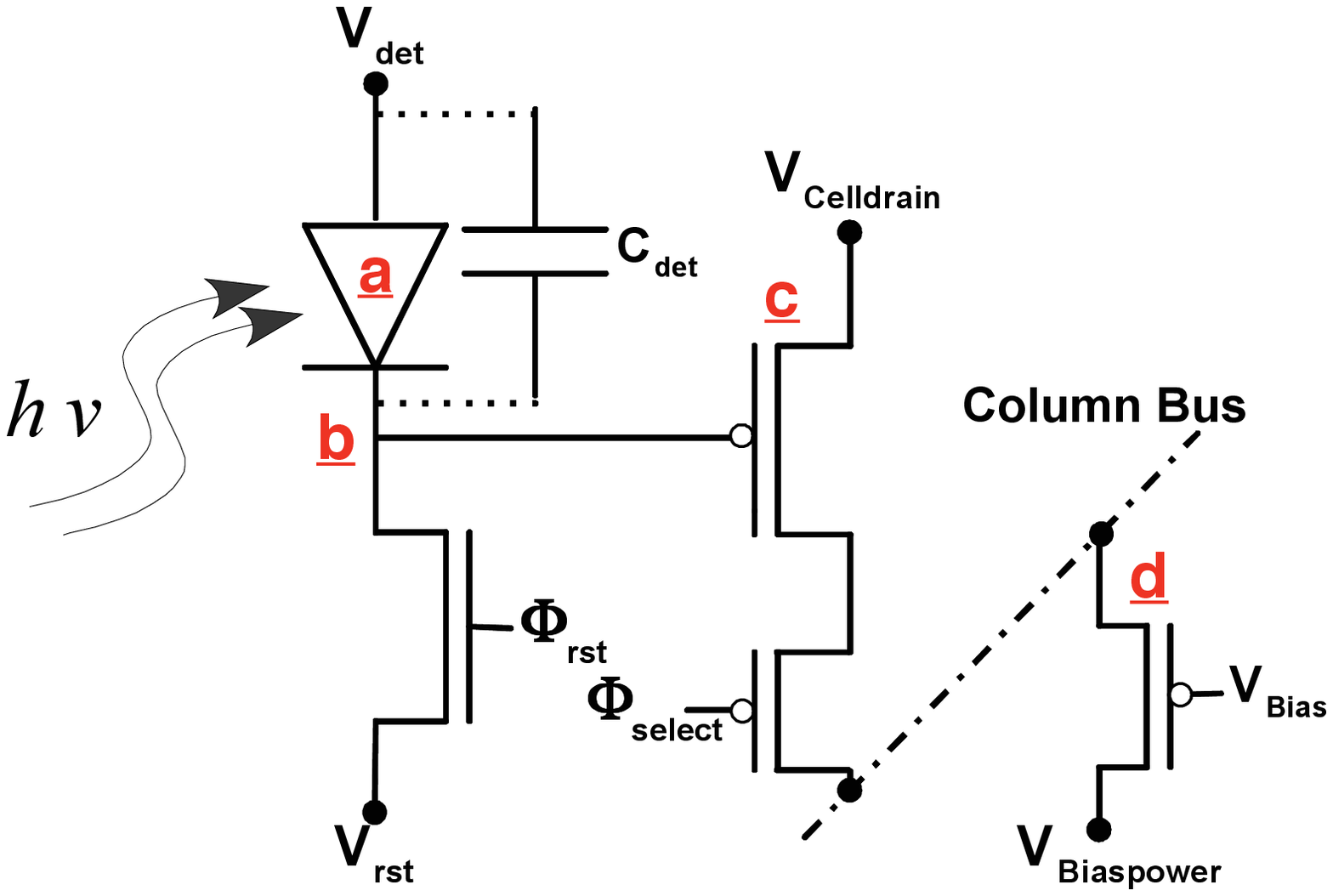}
\caption{This figure shows the H2RG's source follower per detector architecture. The (\underline{a}) photodiode in the HgCdTe detector layer is electrically connected to the ROIC using an indium bump. Although the indium bump is not shown here, functionally it is located at (\underline{b}). The pixel source follower (\underline{c}) is the first FET in the analog signal chain. Its output is multiplexed out on the column bus. ACN is a consequence of the specific way that Teledyne implemented (\underline{d}), the column bus driver. This diagram shows the H2RG's ``unbuffered mode'', in which the column bus is electrically connected to the output. When running in ``buffered mode'', there is an additional output FET that is not shown here. This figure closely follows Fig.~8 of \citet{Loose:2003vh}.\label{fig:schematic}}
\end{center}\end{figure}

Teledyne's H1RG, H2RG, and H4RG detector arrays use a source follower per detector (SFD) architecture. Fig.~\ref{fig:schematic} shows a circuit diagram of an H2RG pixel.\footnote{We cannot legally publish the actual H2RG schematic diagram. Infrared detectors for space are governed by the International Traffic in Arms Limitations (ITAR). The ITAR is a set of United States government regulations that pertain to specified defense-related technologies including \JWST's detectors.} The fundamental noise floor of \JWST's HgCdTe photodiodes is set by kTC noise during pixel reset. Fortunately, kTC noise is completely removed by CDS (or sampling up-the-ramp) and most astronomers never see it. Almost all other noise sources are the result of practical limitations rather than fundamental physics.

The first of these is leakage currents in the HgCdTe photodiodes, \ie dark current. At \JWST's 40~K operating temperature, these are largely modulated by defect states, or traps, in the HgCdTe. Because the HgCdTe will never be perfectly free of defects, there will always be some charge traps and some leakage current. To a first approximation, the noise on integrated leakage current is Poissonian shot noise. For the worst of NIRSpec's two flight SCAs, the median dark current is $<0.01~e^-~s^{-1}$ even when operated at the highest contemplated operating temperature of $\textrm{T}=44.9$~K. The corresponding shot noise per $10.7~s$ CDS integration is about $0.3~e^-$. The noise contribution from leakage current per BSI is about $3~e^-$~rms. Although shot noise on leakage current is clearly important, we did not built it in to NG. We believe that adding in leakage current is best handled at the same time as adding light into models. Sec.~\ref{AddingSources} describes this further and the executable README (see Sec.~\ref{sec:implementation}) includes an example.

After leakage current, the next noise source in the HgCdTe detector layer is Johnson noise in the pixel interconnects. This is a practical rather than fundamental sources because the interconnect resistance is largely determined by design choices at Teledyne. For the NIRSpec H2RGs, which are operated at $T\sim 40$~K, we believe that this contributes $\lesssim 5~e^-$~rms per CDS ($\lesssim 1.8~e^-$~rms per BSI). The interconnect resistance is arguably the last significant noise source in the HgCdTe layer.

The first important noise source in the ROIC is the pixel source follower FET. We expect the pixel source follower to have a $1/f$ noise power spectrum at low frequencies transitioning to white noise at frequencies in the $\sim$few kHz range. Because each pixel has its own source follower, the $1/f$ characteristic is only apparent in noise power spectra made using data from one pixel. The pixel source follower's $1/f$ noise gets converted to white noise in the resulting two dimensional images.

After the pixel source followers, signal voltages are multiplexed to the outputs by the column buses. This is where ACN first appears. ACN is not a fundamental property of the source-follower per detector IR array architecture. Rather, it is caused by a specific design choice in Teledyne's HxRG detector arrays. Working with Teledyne, we traced ACN to a proprietary detail of how the column buses are implemented. Fig.~\ref{fig:schematic} does not show these details. The practical effect is that the even and odd columns on each output each carry some uncorrelated $1/f$ noise that appears in image data as ACN.

After the column buses, the signal is fed either directly to the output in ``unbuffered mode'' or to the output source follower in ``buffered mode''. NIRSpec runs in buffered mode. As such, the output source follower adds additional $1/f$. Unlike the pixel source follower, for which each pixel has its own and the noise appears to be white in image space, $1/f$ noise from the output FET is visible as $1/f$ banding in NG images.

At this point, the signal leaves the H2RG, passes through an interconnect cable, and into the SIDECAR ASIC. From lab testing in the Goddard DCL, we know that the NIRSpec SIDECAR's noise is about $6~e^-$~rms with the inputs shorted.

In addition to affecting the video signal, the SIDECARs also imprint noise through the bias voltages that they provide. In NIRSpec, all H2RG biases are generated by the SIDECARs, and testing in the DCL shows that these are dominated by $1/f$ noise, and moreover are highly sensitive to ASIC temperature. Noise on the biases can imprint itself directly on image data via the DSUB voltage or as picture frame noise by the VBIASGATE and VBIASPOWER bias voltages. \citep{Rauscher:2013ht}. 

Although \JWST NIRSpec provided the impetus for developing NG, the underlying noise model is applicable to all Teledyne HxRG detector systems that we have worked with. By changing the input parameters, NG can simulate different HxRG family detectors, different SIDECARs, and even different controllers than SIDECARs. For example, we are already using NG for trade studies in the \WFIRST Science Definition Team. \WFIRST plans to use H4RG-10 detectors and SIDECARs. They will be configured differently than in \JWST and operated at much warmer temperatures. We have likewise used NG to simulate H2RG and H4RG systems using our laboratory's Gen-III Leach controllers.

\subsection{Simulating Scenes\label{AddingSources}}

In Sec.~\ref{sec:origins}, we briefly mentioned that NG could be used as one of the inputs to simulate astronomical scenes. To simulate a scene that contains astronomical sources, background light, instrument glow, and dark current; one would begin by using NG to generate an ``empty'' noise datacube. The noise cube would contain no charge from any source whatsoever. One would then use separate software to simulate the astronomical scene and effects of the telescope. Finally, as charges are integrated in pixels, they should be summed in to the datacube at the correct pixel locations and arrival times. One would calibrate the resulting datacubes as if they were real astronomical data. For less critical work, one could also generate a scene containing Poissonian shot noise and add it to a 2-dimensional noise image made using NG. The latter approach has the virtue of fast execution time, but it does not accurately model the noise correlations in the time domain. The executable README includes an example that shows how to add shot noise to NG's output.

\section{Software Implementation\label{sec:implementation}}

\begin{figure}[ht]\begin{center}
\includegraphics[width=3in]{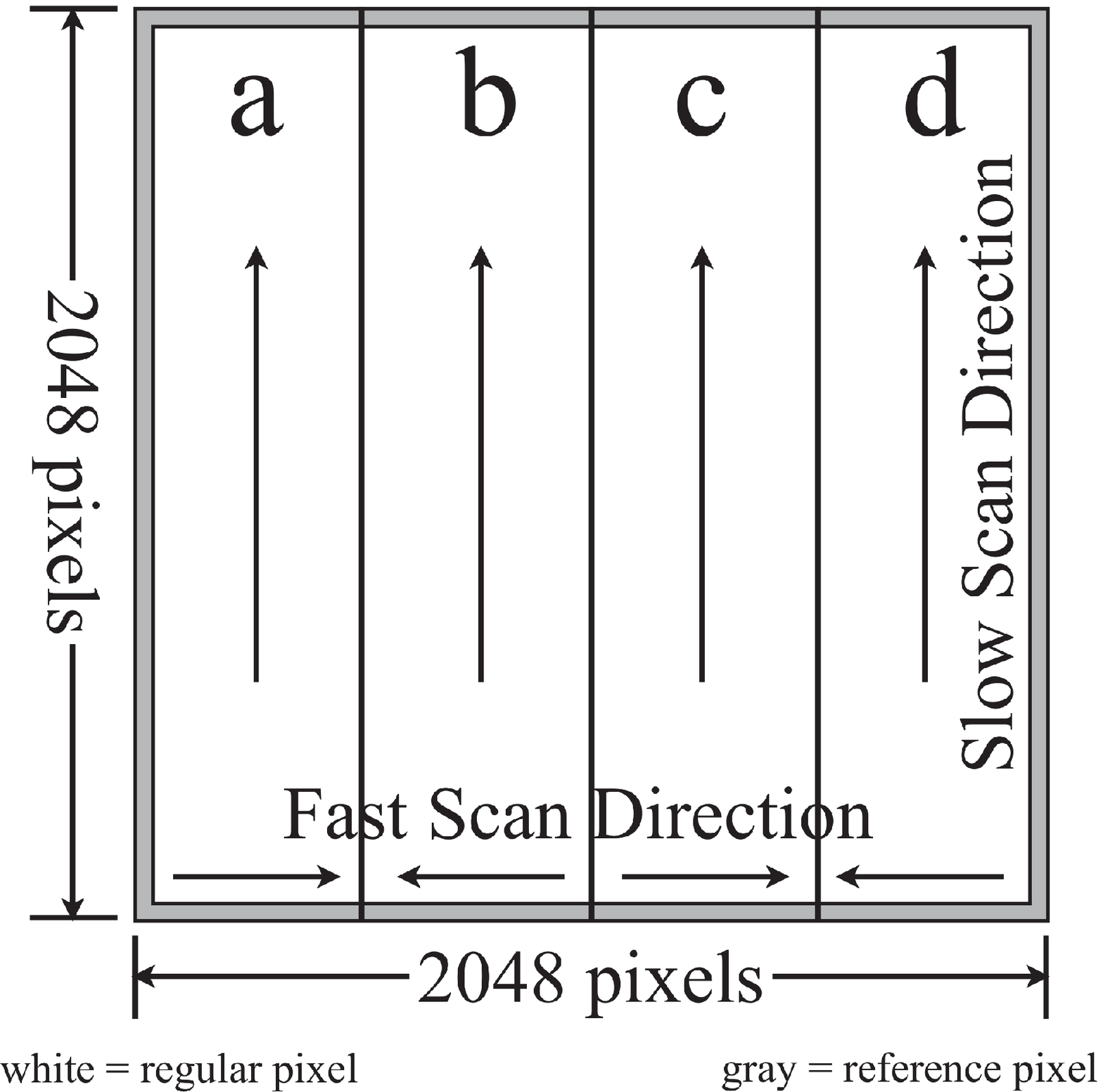}
\caption{The ``traditional'' \JWST H2RG readout pattern clocks pixels sequentially using four video outputs, each running at $100~\textrm{kpix}~s^{-1}$. The \JWST fast and slow scan directions are indicated by arrows. There is a 12~pixel timing overhead at the end of each row for going to the next row and a 1~row overhead at the end of each frame for starting the next frame. \label{fig:readout_pattern}}
\end{center}\end{figure}

By default, NG makes either $2048\times 2048$~pixel H2RG noise images or $n\times 2048\times 2048$~ data cubes. Fig.~\ref{fig:readout_pattern} describes the ``traditional'' \JWST clocking pattern. ``Traditional'' clocking is the same clocking pattern that Teledyne recommends and that is most widely used for astronomy. Depending on the instrument, the number of outputs may differ, as may the duration of end of row and end of frame overheads. When more outputs are used, the same left-to-right, right-to-left, pattern that is shown here continues. The number of outputs, new row overhead, and new frame overhead are user selectable parameters in NG. Tab.~\ref{tab:variables} summarizes all of the user selectable variables and parameters.

\begin{deluxetable}{@{\extracolsep{2pt}}lccl}
\tabletypesize{\scriptsize}
\tablecaption{Instance Variables and Noise Parameters\label{tab:variables}}
\tablewidth{0pt}
\tablehead{\colhead{Name}& \colhead{Default}& \colhead{Unit}& \colhead{Description}}
\startdata
\multicolumn{4}{l}{\underline{HXRGNoise Instance Variables:}}\\
naxis1& 2048& pixel& \parbox[t]{3in}{X-dimension of the FITS cube}\\
naxis2& 2048& pixel& \parbox[t]{3in}{Y-dimension of the FITS cube}\\
naxis3& 1& pixel& \parbox[t]{3in}{Z-dimension of the FITS cube}\\
n\_out& 4& None& \parbox[t]{3in}{Number of detector outputs}\\
nfoh& 1& row& \parbox[t]{3in}{New frame overhead in rows. This allows for a short wait at the end of a frame before starting the next frame.}\\
nroh& 12& pixel& \parbox[t]{3in}{New row overhead in pixels. This allows for a short wait at the end of a row before starting the next row.}\\
dt& $1\times 10^{-5}$& second& \parbox[t]{3in}{Pixel dwell time in seconds}\\
pca0\_file& nirspec\_pca0.fits& None& \parbox[t]{3in}{Name of a FITS file that contains PCA-zero}\\
verbose& False& T$|$F& \parbox[t]{3in}{Set this $=\rm True$ to provide status reporting}\\
\parbox[t]{.75in}{reference\_pixel \_border\_width}\tablenotemark{a}& 4& pixel& \parbox[t]{3in}{Width of reference pixel border around image area}\\
\parbox[t]{1in}{reverse\_scan \_direction}& False& T$|$F& \parbox[t]{3in}{Set this $=\rm True$ to reverse the fast scanner readout directions. This capability was added to support Teledyne's programmable fast scan readout directions. The default setting $=\rm False$ corresponds to what HxRG detectors default to upon power up.}\\ \\
\multicolumn{4}{l}{\underline{HXRGNoise.mknoise Method Parameters:}}\\
o\_file& None& None&\parbox[t]{3in}{Output filename}\\
pedestal& 4&$e^-$& \parbox[t]{3in}{Magnitude of pedestal drift}\\
rd\_noise& 5.2&$e^-$& \parbox[t]{3in}{Standard deviation of read noise}\\
c\_pink& 3&$e^-$& \parbox[t]{3in}{Standard deviation of correlated pink noise}\\
u\_pink& 1&$e^-$& \parbox[t]{3in}{Standard deviation of uncorrelated pink noise}\\
acn& 0.5&$e^-$& \parbox[t]{3in}{Standard deviation of alterating column noise}\\
pca0& 0.2&$e^-$& \parbox[t]{3in}{Standard deviation of pca0}\\
\parbox[t]{1in}{reference\_pixel \_noise\_ratio}& 0.8&None& \parbox[t]{3in}{Ratio of the standard deviation of the reference pixels to the regular pixels. Reference pixels are usually a little lower noise.}\\
ktc\_noise& 29& $e^-$& \parbox[t]{3in}{Set this equal to $\sqrt{k T C}/q_e$, where $k$ is Boltzmann's constant, $T$ is detector operating temperature, $C$ is pixel capacitance, and $q_e$ is the fundamental charge. This default value is appropriate for an H2RG. H4RG series detectors will likely have a lower value.}\\
bias\_offset& 5000& $e^-$& \parbox[t]{3in}{In datacubes, pixels start integrating at about this level.}\\
bias\_amp& 500& None& \parbox[t]{3in}{This is used to approximate the bias pattern that is seen in individual HxRG reads. To approximate the bias pattern, we multiply the contents of pca0\_file by this factor.}
\enddata
\tablenotetext{a}{To save space, long names wrap to the next line with a leading underscore. When written without wrapping, this variable becomes reference\_pixel\_border\_width. Other variable names that wrap include reverse\_scan\_direction and reference\_pixel\_noise\_ratio.}
\end{deluxetable}

The HxRG architecture is very flexible, incorporating reversible scan directions and programmable subarrays. NG supports changing the fast scan readout directions and basic subarrays.\footnote{NG does not currently support the more complicated situation of interleaved subarrays and science data. One of our aims in releasing the source code is to enable others to add this capability which is often instrument specific.} NG defaults to the same fast scan directions that Teledyne HxRG detectors default to upon power up. We believe that this is appropriate for most astronomical cameras. To reverse the fast scanners, set the reverse\_scan\_direction instance variable to True when instantiating an HXRGNoise object. We have not implemented reversible slow scanners in NG because the same result can be accomplished using python slices on the output from NG. The README file provides an example that shows how to reverse both the fast and slow scanners.

\begin{figure}[ht]\begin{center}
\includegraphics[width=6in]{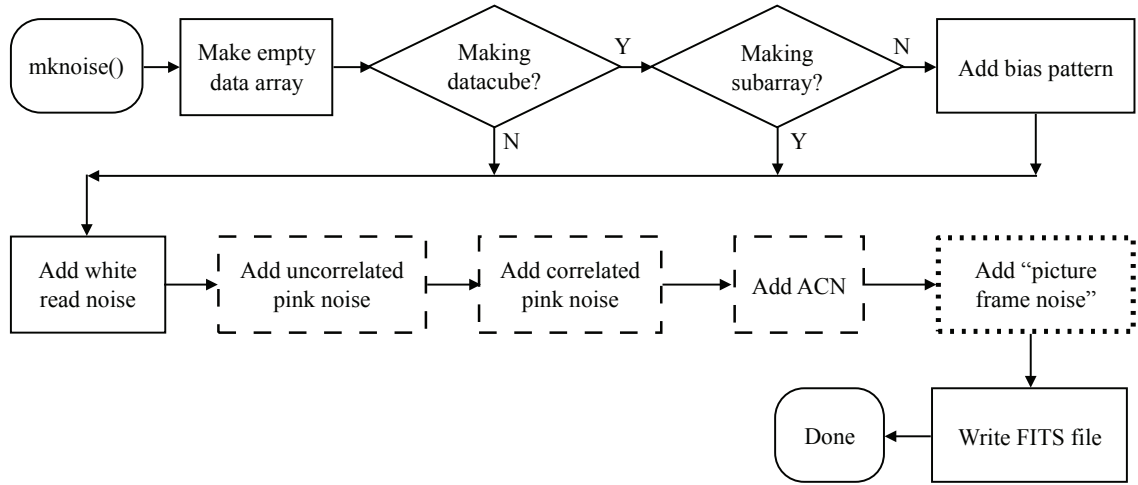}
\caption{This figure shows the execution flow though the mknoise() method. This is the top level method for generating noise. Boxes with solid borders work in the time domain. Boxes with long dashed borders work primarily in Fourier space. Picture frame noise, dotted border, exploits both Fourier space and the time domain to add a time-modulated image of the zeroth principal noise component, PCA-zero.\label{fig:flowchart}}
\end{center}\end{figure}

It is best to refer to the examples and source code when trying to understand the implementation. The file 01\_README.ipynb is an executable IPython Notebook that explains NG in more detail, and provides a few examples. Fig.~\ref{fig:flowchart} shows a flow diagram for the mknoise() method. This shows the sequence in which different noise components are added, indicating which steps are done in the time domain and which are done in Fourier space. In the following paragraphs, we explain a few aspects of NG in more detail than is practical in the code comments.

\subsection{Generating Non-Periodic Pink Noise\label{sec:ffts}}

We make pink noise by filtering white noise in Fourier space. Because the Fourier transform assumes that its input data are periodic, care must be taken to avoid generating periodic pink noise. Although periodic pink noise still has a $1/f$ power spectrum, real physical systems usually do not drift back to their starting points at the ends of experiments.

NG uses the HXRGNoise class's pink\_noise() method to generate non-periodic pink noise. We generate $2\times$ as many Gauss-random numbers as are ultimately required. We compute the FFT of this seed vector and multiply it by a pinkening filter. Finally, we invert the FFT and keep only the first half of the data. This procedure produces $1/f$-noise that randomly walks away from the initial value as desired rather than wrapping back around on itself as would otherwise be the case. Finally, we offset and scale the data to have the same mean and standard deviation as the initial random seed vector. For long seed vectors, pink\_noise() produces noise that is very close to pink and that drifts away from the starting point as desired.

NG can generate both correlated and uncorrelated pink noise. Here correlated means that the same pink noise appears on all outputs. Uncorrelated means only that each output can have some pink noise that is unique to itself.

\subsection{Generating ACN\label{sec:acn}}

Observationally, ACN manifests as a feature at the Nyquist frequency with a $1/f$-like shoulder on the low frequency side. Physically, we believe that the alternating column pattern at the Nyquist frequency is likely carrying much of the same $1/f$ that appears at low frequency, but up-converted to the Nyquist frequency. To simulate this, we generate non-periodic pink noise as described in Sec.~\ref{sec:ffts} for the even and odd columns of each output separately. For example, NIRSpec's H2RGs are read out using four outputs. To simulate ACN, we use the pink\_noise() method to generate eight vectors of pink noise.

\subsection{Generating Picture Frame Noise\label{sec:pf}}

PCA of the NIRSpec detector subsystem showed that there is a picture frame component that drifts in and out during integrations. Other applications that do not use SIDECARs, or that run their SIDECARs and/or detectors warmer, may not see the picture frame as it is described here. To completely turn off the picture frame component in NG, set $\textrm{pca0\_amp}=0$ when calling HXRGNoise's mknoise() method.

Extensive testing in the DCL showed that the appearance of the picture frame in any one NIRSpec integration can be approximated by $1/f$ noise modulating 0$^\textrm{th}$ PCA component, which is the dominant one.

\section{Comparing Real and Simulated NIRSpec Data\label{sec:comparison}}

\begin{figure}[ht]\begin{center}
\includegraphics[width=6in]{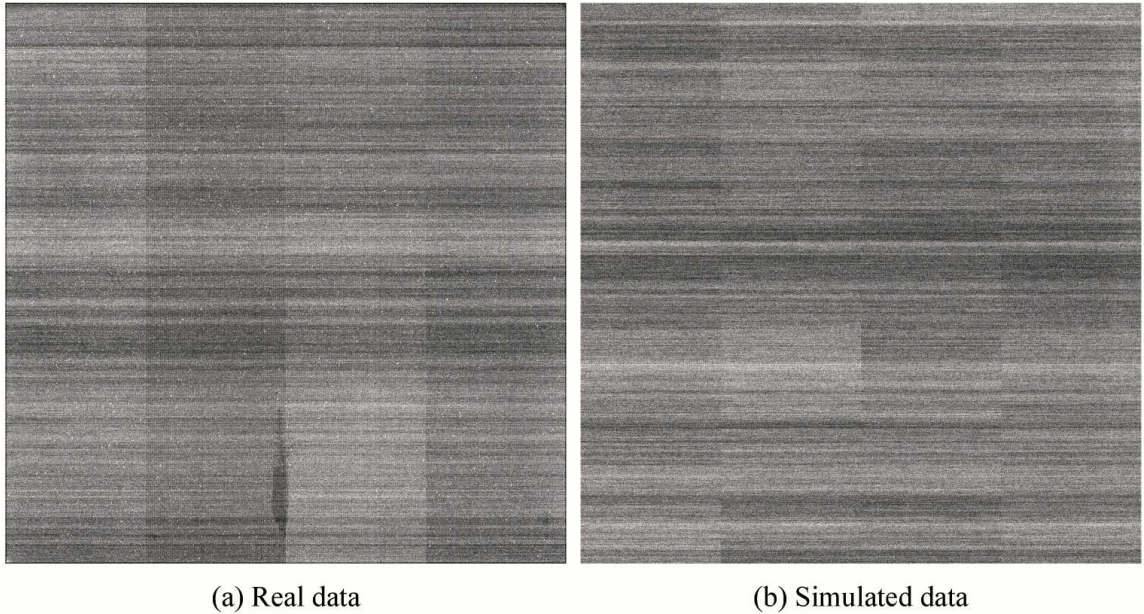}
\caption{This figure compares: (a) real and (b) simulated \JWST NIRSpec slope images. The read noise is about $6~e^-$~rms in both images. The integration time, $=934~\textrm{seconds}$, is the same for both. Both are on the same approximately $\left(-10,+10 \right)~e^-$ grayscale after dark subtraction. The reference pixels in the real data appear dark because the regular pixels contain integrated dark current whereas those in the simulated data do not. Both show significant white, $1/f$, and ACN contributions. NIRSpec's four video outputs are clearly seen. The picture frame is not obvious in either image, although is some integrations it can become more obvious. The elongated dark feature near the bottom-center in the real data is an epoxy void.\label{fig:real_and_simulated}}
\end{center}\end{figure}

Fig.~\ref{fig:real_and_simulated} compares real and simulated NIRSpec after calibrating the 88-frame up-the-ramp sampled cube to a reference pixel corrected image. Here we compare a two dimensional simulation to the calibrated image. NG could also be used to produce an 88-frame cube which could be calibrated. In practice, we seldom do this because the execution time is very long. However, we have verified that it works.

The real detector has an epoxy void and hot pixels. The void appears as a dark feature near the bottom-center. Teledyne uses an epoxy backfill between the silicon ROIC and HgCdTe detector layer. Among other things, this increases mechanical strength. Sometimes, however, the backfill process is not perfect and voids appear where the epoxy fails to flow. In NG, we do not attempt to simulate voids because each one is unique.

Visibly, the simulated image is a reasonable facsimile of the real data. This is not surprising since NIRSpec's noise is close to stationary and the input parameters were tuned to match the real system in Fourier space. NIRSpec's four video outputs are clearly seen. The horizontal banding is caused by $1/f$ noise. Upon close inspection, ACN can be seen in both and picture frame artifacts may be detectable along the right hand edge. The examples include FITS files that can be examined using a FITS viewer such as DS9 to better see these fine details.

\begin{figure}[ht]\begin{center}
\includegraphics[height=6.75in]{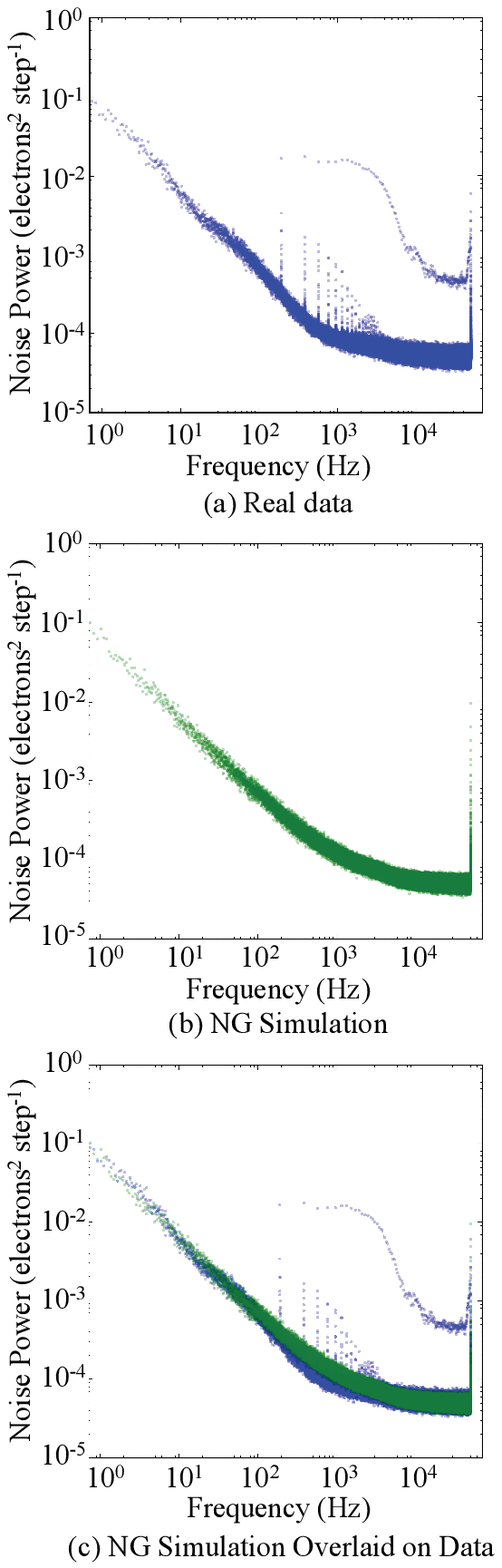}
\caption{This figure provides a more quantitative comparison between real and simulated images. (a) Shows the noise power spectrum of the actual NIRSpec flight detector, 491, while (b) is the power spectrum of a simulated image using the default noise parameters in mknoise(). Finally, (c) overlays the simulated data on the real data. NG simulates the $1/f$ characteristic at low frequency, ACN at 50~kHz, and white noise at intermediate frequencies. Much of the ``fuzz'' that appears in the real data is due to interpolation over bad pixels. There are no bad pixels in the simulated data.\label{fig:real_and_simulated_ps}}
\end{center}\end{figure}

Fig.~\ref{fig:real_and_simulated_ps} provides a quantitative comparison in Fourier space. NG captures the main features including; $1/f$ at low frequency, white noise at mid frequencies, and ACN at 50~kHz. Much of the ``fuzz'' that is present in the real power spectrum is an artifact from interpolating over bad pixels. NG does not simulate bad pixels. Near the ``knee'', where $1/f$ transitions to white noise in the real data,
there is some structure that NG does not model. If this were to become important, one solution would be to allow the user to explicitly specify a pinkening filter (\ie frequency dependent weights) on the command line.

\section{Summary}\label{sec:summary}

This paper describes NG, the noise generator that we wrote for the \JWST NIRSpec detector subsystem. NG is capable of simulating both stationary and non-stationary read noise components. Compared to more basic simulators that include white and $1/f$ noise; it builds in the correct Teledyne H2RG clocking pattern and timing and includes ACN and picture frame noise. With appropriate choice of input parameters, NG can simulate other HxRG family Teledyne detectors including the H4RG and H1RG.

\begin{acknowledgments}
This work was supported by NASA as part of the \JWST Project. NG would not have been possible without many invaluable contributions from the NIRSpec Detector Subsystem Team at Goddard, the \JWST ISIM team, and the ESA NIRSpec Team. With regard strictly to minimizing and understanding the as-built system noise, Yiting Wen and Markus Loose played the lead roles in tuning the system for peak performance. More than anybody else, Harvey Moseley realized that understanding the noise in detail would allow us to develop better tools for removing it. Dale Fixsen provided the mathematical tools to do so, and Rick Arendt implemented them. I wish to thank Brent Mott for his superb technical management of the NIRSpec Detector Subsystem throughout the challenging integration and test campaign, and also Donna Wilson for taking ownership of the SIDECAR ASIC and mastering the practicalities of making it work for \JWST. Ray Wright routinely performed above and beyond the call of duty by learning Python at about the same time as I did and developing tools for analyzing \JWST telemetry that greatly facilitated understanding what causes the noise. Pierre Ferruit of ESA provided many helpful comments on both the source code and manuscript. Chaz Shapiro of NASA JPL steered me toward better ways of handling several challenges in the Python implementation. I wish to thank the referee for providing many helpful comments, on both the text and source code. A great many other people contributed important things, but I will stop here lest this acknowledgment become so long that nobody reads it.
\end{acknowledgments}


\begin{thebibliography}{} 
\expandafter\ifx\csname natexlab\endcsname\relax\def\natexlab#1{#1}\fi 

\bibitem[{Bacon {et~al.}(2004)Bacon, McMurtry, Pipher, Forrest, Garnett, \&
  Lee}]{Bacon:2004dq}
Bacon, C., McMurtry, C.~W., Pipher, J.~L., {et~al.} 2004, Proc SPIE, 5563, 35

\bibitem[{Bacon {et~al.}(2005)Bacon, McMurtry, Pipher, Forrest, \&
  Garnett}]{Bacon:2005bl}
Bacon, C.~M., McMurtry, C.~W., Pipher, J.~L., Forrest, W.~J., \& Garnett, J.~D.
  2005, Proc SPIE, 5902, 116

\bibitem[{Loose {et~al.}(2003)Loose, Farris, Garnett, Hall, \&
  Kozlowski}]{Loose:2003vh}
Loose, M., Farris, M.~C., Garnett, J.~D., Hall, D. N.~B., \& Kozlowski, L.~J.
  2003, Proc SPIE, 4850, 867

\bibitem[{Moseley {et~al.}(2010)Moseley, Arendt, Fixsen, Lindler, Loose, \&
  Rauscher}]{Moseley:2010kc}
Moseley, S.~H., Arendt, R.~G., Fixsen, D.~J., {et~al.} 2010, Proc SPIE, 7742,
  36

\bibitem[{Rauscher {et~al.}(2004)Rauscher, Figer, Regan, Boeker, Garnett, Hill,
  Bagnasco, Balleza, Barney, Bergeron, Brambora, Connelly, Derro, DiPirro,
  Doria-Warner, Ericsson, Glazer, Greene, Hall, Jacobson, Jakobsen, Johnson,
  Johnson, Krebs, Krebs, Lambros, Likins, Manthripragada, Martineau, Morse,
  Moseley, Mott, Muench, Park, Parker, Polidan, Rashford, Shakoorzadeh, Sharma,
  Strada, Waczynski, Wen, Wong, Yagelowich, \& Zuray}]{Rauscher:2004ga}
Rauscher, B.~J., Figer, D.~F., Regan, M.~W., {et~al.} 2004, Proc SPIE, 5487,
  710

\bibitem[{Rauscher {et~al.}(2013)Rauscher, Arendt, Fixsen, Greenhouse, Lander,
  Lindler, Loose, Moseley, Mott, Wen, Wilson, \& Xenophontos}]{Rauscher:2013ht}
Rauscher, B.~J., Arendt, R.~G., Fixsen, D.~J., {et~al.} 2013, in Proc SPIE, ed.
  H.~A. MacEwen \& J.~B. Breckinridge (SPIE), 886005

\bibitem[{Rauscher {et~al.}(2014)Rauscher, Boehm, Cagiano, Delo, Foltz,
  Greenhouse, Hickey, Hill, Kan, Lindler, Mott, Waczynski, \&
  Wen}]{Rauscher:2014wk}
Rauscher, B.~J., Boehm, N., Cagiano, S., {et~al.} 2014, PASP, 126, 739

\end{thebibliography}
\end{document}